\documentclass[prl,twocolumn,showpacs]{revtex4}

\usepackage{graphicx}
\usepackage{amsmath}

\newcommand{\parref}[1]{(\ref{#1})}

\newcommand{\tento}[1]{10^{#1}}
\newcommand{\vect}[1]{\mathbf{#1}}
\newcommand{\units}[1]{\,\mathrm{#1}}

\newcommand{\angs}{\textrm{\AA}}

\newcommand{\dexp}[1]{\cdot 10^{#1}}

\newcommand{\Otwo}{\ensuremath{\mathrm{O}_2}}
\newcommand{\Ntwo}{\ensuremath{\mathrm{N}_2}}

\newcommand{\cwi}{
CWI, P.O. Box 94079, 1090 GB Amsterdam, The Netherlands
}

\begin{document}

\title{Interaction of streamers in air and other oxygen--nitrogen mixtures}
\author{A. Luque$^1$, U. Ebert$^{1,2}$, and W. Hundsdorfer$^1$}
\affiliation{$^1$\cwi,\\$^2$ Dept.\ Appl.\ Physics, Eindhoven Univ.\ Techn., The Netherlands}

\date{\today}

\begin{abstract}
The interaction of streamers in nitrogen-oxygen mixtures such as air is studied. First, an efficient
method for fully three-dimensional streamer simulations in multiprocessor machines is introduced.
With its help, we find two competing mechanisms how two adjacent streamers can interact: through
electrostatic repulsion and through attraction due to nonlocal photo-ionization. The non-intuitive effects
of pressure and of the nitrogen-oxygen ratio are discussed. As photo-ionization is experimentally difficult
to access, we finally suggest to measure it indirectly through streamer interactions. 
\end{abstract}

\maketitle

Streamer discharges are fundamental building blocks of sparks and lightning in any ionizable matter; they
are thin plasma channels that penetrate nonconducting media suddenly exposed to an intense electric
field. They propagate by enhancing the electric field at their tip to a level that facilitates an ionization
reaction by electron impact \cite{Raizer,PSST06}.
Streamers are also the mechanism underlying sprites \cite{PaskoGRL98, PaskoBook, NielsenGRL07}; these
are large atmospheric discharges above thunderclouds that, despite being tens of
kilometers wide and intensely luminous, were not reported until 1990 \cite{fra1990}.  Although the investigation of streamers concentrates
mainly in gaseous media,
they have also been studied in dense matter, such as semiconductors
\cite{Rodin2004} and oil \cite{Massala01}.  Streamers have also received 
attention in the context of Laplacian-driven growth dynamics \cite{arr2002} 
and a strong analogy with viscous fingering, in particular
Hele-Shaw flows \cite{meu2005}, has been established.

Both in laboratory \cite{Briels2006} and in nature \cite{ger2000}, streamers appear frequently in trees or
bundles. As their heads carry a substantial net electrical charge of equal polarity that creates the local field
enhancement, they clearly must repel each other electrostatically which probably causes the ``carrot''-like
conical shape of sprites. On the other hand, recent sprite observations~\cite{Cummer06} as well as
streamer experiments (\cite{Winands06}, Fig.~7, \cite{Briels2006}, Fig.~6) 
also show the opposite: streamers attract each other and coalesce.

\begin{figure}
\includegraphics[width=0.4 \textwidth]{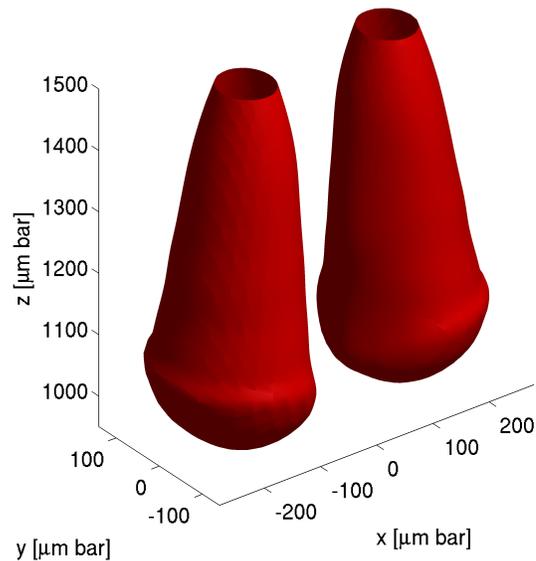}
\caption{\label{isoelectrons}
  Two negative streamers in nitrogen at atmospheric pressure advancing downwards and repelling each other;
  shown are surfaces of constant electron density in an advanced state of evolution within a constant
  background field. See the text for details.}
\end{figure}

Up to now, streamer interactions have not been studied much theoretically, and streamer attraction has not been
predicted at all. In coarse grained phenomenological
models for a streamer tree as a whole~\cite{Akyuz03}, the repulsive electrostatic interaction between streamers
is taken into account. In a more microscopic, but still largely simplified model, Naidis~\cite{Naidis96}
studied the corrections to the streamer velocity due to electrostatic interaction with neighboring streamers.
In \cite{FingerLetter}, two authors of the present letter have studied a microscopic ``fluid'' model for a
periodic array of negative streamers in two spatial dimensions, where they show that shape, velocity and
electrodynamics of an array of streamers substantially differs from those of single streamers due to their
electrostatic interaction, but attraction or repulsion were excluded by the approach. 
Due to the difficulty to represent this multiscale process \cite{PSST06} in a numerically efficient manner,
 only recently it has become possible to simulate streamers in full 3D \cite{Kulikovsky98b, Pancheshnyi05b}. 
We here present a numerical method to handle this problem, and we apply it to the interaction
of streamers in complex gases like air where a nonlocal photon mediated ionization reaction has to be taken
into account. We find that when varying gas composition and pressure, streamers can either repel or attract each
other. The transition occurs in an unexpected manner, and is not simply determined by an ionization length.

The photon mediated ionization reaction in air and other nitrogen oxygen mixtures is experimentally not
easily accessible, but forms a basic ingredient of the present theory of streamers in air~\cite{Pancheshnyi05b,
LiuPasko06, Segur06, Bourdon07, Luque07}; all these simulations are based on the single experimental
measurement of Penney and Hummert in 1970~\cite{pen1970}. Our theoretical results suggest that this reaction
could be deduced from experiments on coalescence or repulsion of adjacent streamers as a function of pressure
and gas composition.

We study the minimal streamer model  \cite{vit1994} extended by the nonlocal photo-ionization
reaction characteristic for nitrogen oxygen mixtures like air. It consists of continuity equations for
electron and ion densities $n_{e,+}$ coupled to the electrical field $\vect{E}$ that is determined 
by the potential on the outer boundaries and space charge effects
\begin{eqnarray}
  \label{fluid}
  \partial_t n_e & = &
   \nabla \cdot (n_e \mu_e \vect{E}) + D_e \nabla^2 n_e + S_i + S_{ph},
     \label{sigma}\\
  \partial_t n_+ & = & S_i + S_{ph}, \label{rho}\\
\epsilon_0 \nabla\cdot\vect{E} &=& {\rm e} (n_+ - n_-), ~~~\vect{E}=-\nabla\phi. \label{field}
\end{eqnarray}
Here $\mu_e$ is the electron mobility, $D_e$ is the electron diffusion coefficient and $e$ is the elementary
charge. Ion mobility, much smaller than electron mobility, is neglected. To fully focus
on the influence of photo-ionization, electron attachment on oxygen is here 
neglected as well, and we use transport parameters for pure nitrogen as in previous work~\cite{PSST06,Montijn06}. 
The source terms for additional electron-ion pairs are the local impact ionization $S_i$ in Townsend approximation,
$S_i = n_e \mu_e |\vect{E}| \alpha(|{\bf E}|) = n_e \mu_e|\vect{E}|
    \alpha_0 e^{-E_0/|{\bf E}|}$,
where $\alpha_0$ is the ionization cross section and $E_0$ is the threshold field, and the nonlocal 
photo-ionization according to the model for oxygen-nitrogen mixtures developed by Zhelezniak 
{\it et al.}~\cite{zhe1982} 
\begin{equation}
  \newcommand{\R}{\vect{r}}
  \label{Sph_int_dim}
  S_{ph}(\vect{\R}) = \frac{\xi A(p)}{4\pi} \int
  \frac{h(p|\vect{\R} - \vect{\R}'|)
           S_i(\vect{\R}')d^3(p\vect{\R}')}
       {|p\vect{\R} - p\vect{\R}'|^2},
\end{equation}
with $A(p)=p_q/(p+p_q)$.  Here it is assumed that accelerated 
electrons excite the $b^1\Pi_u$,
$b'^1\Sigma_u^+$ and $c'^1_4\Sigma_u^+$ states of nitrogen
by impact with a rate $\xi S_i$ where $S_i$ is the local impact ionization rate
and $\xi$ a proportionality factor.
These nitrogen states can deexcite under emission of a photon in the wavelength range
$980-1025 \,\angs$ that can ionize oxygen molecules \cite{zhe1982, pen1970}. The absorption length
of these photons by oxygen is obviously inversely proportional to the oxygen 
partial pressure $p_{O_2}$.
Introducing the oxygen concentration in the gas as $\eta=p_{O_2}/p$, 
the absorption function of photoionizing radiation $h(p r)$ is characterized by the two length scales 
$pr_{min} \approx 380\; \mu{\rm m \cdot bar}/\eta$ and $pr_{max} \approx 6.6\; \mu{\rm m \cdot bar}/\eta$.
Above a critical gas pressure that we take as $p_q=60 \units{Torr}=80 \units{mbar}$~\cite{Luque07},
the excited nitrogen states can be quenched by collisions with neutrals, hence suppressing photo-emission; 
this is taken into account through the prefactor $A(p)$.

A major difficulty in the theoretical study of streamer-streamer interactions is the development of an
efficient numerical code for streamer simulations with their inner multiscale structure in three dimensions.
Here we present a locally adaptive and parallelizable 
approach to this problem. Our method has in common with the one described
in~\cite{Kulikovsky98b} that cylindrical coordinates and a uniform grid in the angular dimension
are used, but in the $r, z$-projection we apply the grid refinement scheme of \cite{Montijn06}
to resolve better the thin, pancake-like shape of the space-charge layer.  Furthermore,
to allow the parallel solution of the Poisson equation
in multi-processor machines we perform a Fourier transformation in the
angular coordinate $\theta$,
$  \tilde{\phi}_{k}(r, z) = \sum_{n = 0}^{N - 1} \phi (r, z, \theta_n)
       e^{-ik\theta_n},$
where $N$ is the number of grid cells in the $\theta$ direction and
$\theta_n = 2\pi n/ N$. For each separate mode $k$, a two-dimensional
Helmholtz equation has to be solved:
\begin{equation}
  \nabla_{rz}^2 \tilde{\phi}_k
     + \frac{|w_k|^2}{r^2} \tilde{\phi}_k = -\frac{e}{\epsilon_0}
     (\tilde{n}_{+k} - \tilde{n}_{ek}),
     \label{helmholtz}
\end{equation}
where a tilde $\tilde{}$
represents the Fourier transform of a quantity and $|w_k|^2 =
\frac{2}{\Delta \theta^2} (1 - \cos k\Delta \theta)$.
For each Fourier mode, we apply the refinement algorithm described in
\cite{wac2005}, which is trivially generalized to solve the Helmholtz
instead of the Poisson equation.  The advantage of solving the
electrostatic part of the problem in the Fourier domain is that each
of the Fourier modes is decoupled from the rest and hence it can be
solved in parallel in a multi-processor computer.
Once the electric field is calculated in Fourier space, the field in real space
is derived through an inverse Fast Fourier Transform; and the convection-diffusion-reaction system
\parref{fluid}-\parref{rho} is integrated as detailed in \cite{Montijn06}.
The photo-ionization term is computed in a Helmholtz PDE approach as described in \cite{Luque07},
and in the angular direction with a scheme of Fourier transformations and parallel solving that is completely
equivalent to the one applied to the Poisson equation.

\begin{figure}
\includegraphics[width=0.5 \textwidth]{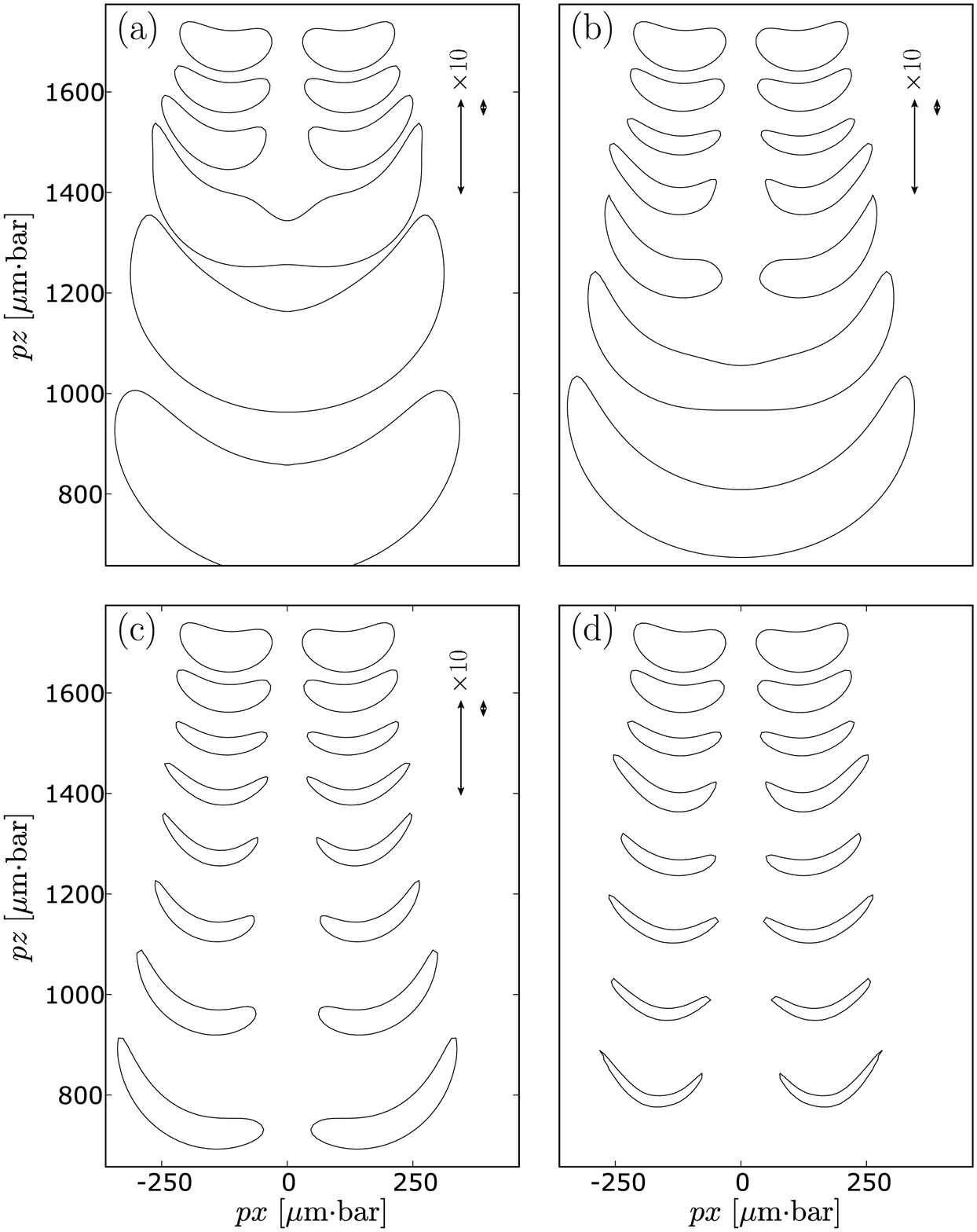}
\caption{\label{pressures}
  Evolution of the space charge layers of two adjacent negative streamers at different pressures in air. 
(a) $p=0.07 {\rm mbar}$,
  (b) $p=1 \units{bar}$, (c) $p=50 \units{bar}$, (d) $p \to \infty$. (a) corresponds to $A(p)\approx1$ and (d) 
to $A(p)=0$, which is equivalent to pure nitrogen since there is no photo-ionization (even though in pure nitrogen the length-scales formally diverge).
  The axes show reduced lengths $p\cdot x$, $p\cdot z$ to exhibit similarities. In the upper right
  corners, the two photo-ionization lengths are inserted as vertical bars.  A multiplicative factor is used where the lengths are too long to fit into the figure.}
\end{figure}

We now use our model and numerical algorithm to study the interaction between two streamers.  
We focus on the influence of $(i)$ the pressure in air and of $(ii)$ the oxygen-nitrogen 
ratio at standard pressure; standard temperature is always assumed.  $(i)$
The comparison of streamers at different pressures $p$ relies on scaling lengths, times etc. with appropriate powers 
of $p$. These similarity laws are strictly valid 
in the minimal streamer model~\cite{PSST06,Montijn06}; they are broken by photo-ionization 
\parref{Sph_int_dim} when the pressure reaches the quenching pressure $p_q$; for $p\gtrsim p_q$, photo-ionization
at unchanged gas composition is increasingly suppressed like $A(p)$ \cite{LiuPasko06}. Similarity or Townsend scaling
is further discussed in Sect.~1.2 of~\cite{TanjaSca08}. $(ii)$ The O$_2$:N$_2$ ratio changes 
the local photoionization rate which is proportional to it,
while the absorption lengths are inversely proportional to it.

First, negative streamers are investigated, because they propagate even in the absence 
of photo-ionization. Then it is shown that positive streamers behave similarly. In all simulations, we use two 
identical Gaussian seeds of reduced width (according to Townsend scaling)
$p\cdot w=73.6 \units{\mu m \cdot bar}$ 
and amplitude $p^{-2} \cdot n_{e,max}=1.4 \dexp{-2} 
\units{\mu m^{-3} \cdot bar^{-2}}$ 
separated by a reduced distance of $p\cdot d=230 \mu {\rm m\cdot bar}$ as an initial condition.
They are exposed to a homogeneous and constant background
electric field $E_b/p = 80 \units{kV/(cm\;bar)}$ in the positive $z$ direction.
The streamers on reduced length, time and density scales are similar
if photo-ionization is neglected.

The simulations were performed with an angular resolution of $\Delta \theta = 2\pi / 64$ ($N = 64$), 
but we also checked that the results are stable when we double the number of angular grid cells. 
Figure~1 shows a surface of equal electron density for the discharge in nitrogen at normal pressure and
temperature at time $t=1.56 \units{ns}$. 
Figures~2-4 show the space charge layers at the streamer heads 
(more precisely, the half maximum line at the respective 
time) in the plane
intersecting the two streamer axes in timesteps of $0.12 \units{ns \cdot bar}$;
the first snapshot is at time $0.84 \units{ns \cdot bar}$.

Figure 2 demonstrates the effect of pressure change for similar streamers in artificial air, i.e., 
in an oxygen nitrogen mixture of ratio 20:80, corresponding to $\eta=0.2$. 
The panels show (a) air at a pressure of $0.05\units{torr} \approx 0.07 \units{mbar}$, which is the 
atmospheric pressure at approximately $70\units{km}$ height, where sprites are frequently observed,
(b) air at 1 bar, (c) air at 50 bar, and (d) pure nitrogen which in our model is equivalent to air 
at infinite pressure.

\begin{figure}
\includegraphics[width=0.5 \textwidth]{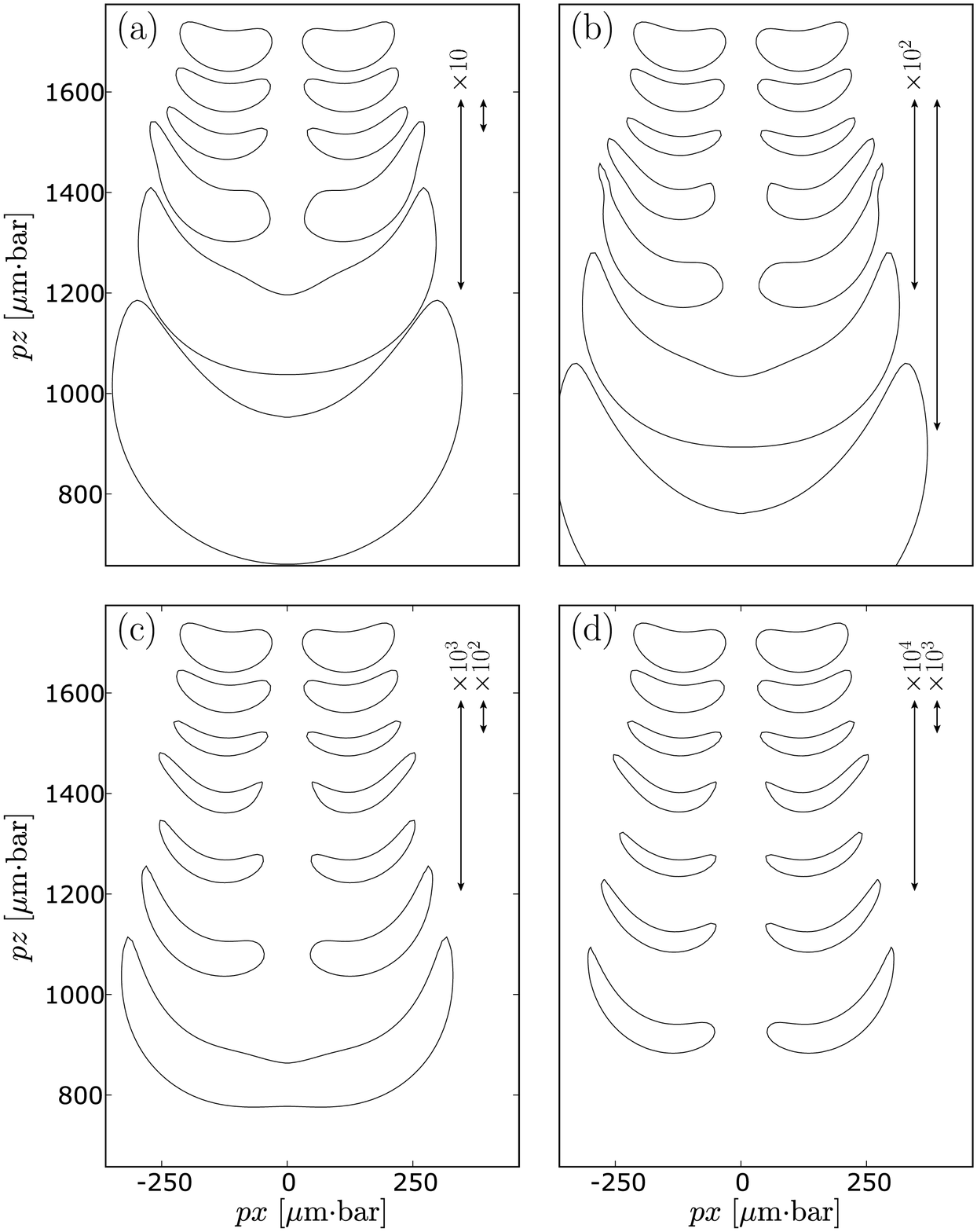}
\caption{\label{gammas}
  Evolution of the space charge layers of two adjacent streamers at atmospheric pressure, but different
  concentrations of \Otwo: $\eta=p_{O_2}/p = \tento{-1},~\tento{-2},~\tento{-3},~\tento{-4}$ in panels 
  (a)--(d). The two photo-ionization lengths are again inserted as vertical bars in the upper right corners.}
\end{figure}

The first observation is that two streamers do not always repel each other. 
Rather there are two qualitatively different regimes: at a pressure of 50 bar and above,
when the fast collisional quenching of \Ntwo\  molecules suppresses photoionization, 
the streamers repel each other electrostatically due to the net negative charge in their heads.
However, at atmospheric pressure and below, the streamers attract each other:
a cloud of electrons is created between the two streamers which eventually makes them coalesce 
into a single, wider one. As it has been suggested that the photo-ionization length could determine
length scales in the streamer head \cite{kul2000-2}, the two reduced photo-ionization lengths are  
indicated by the vertical bars in the upper right corners, they are the same in all panels. 
Obviously, there is no simple relation between these lengths and the observed repulsion or attraction;
the interaction is rather governed by the quenching prefactor $A(p)$.

Figure 3 demonstrates the effect of the oxygen concentration. For fixed atmospheric pressure,
the relative oxygen concentration $\eta=p_{O_2}/p$ is reduced to $\tento{-1},~\tento{-2},~\tento{-3},
~\tento{-4}$ in panels (a)--(d). As the photo-ionization lengths scale inversely with the oxygen 
concentration, these lengths increase by factors 10 from one panel to the next while the prefactor 
decreases with a factor $\tento{-1}$. The attraction between the streamers decreases with
decreasing oxygen concentration until it is not visible anymore in panel (d) for $\eta=\tento{-4}$.

\begin{figure}
\includegraphics[width=0.5 \textwidth]{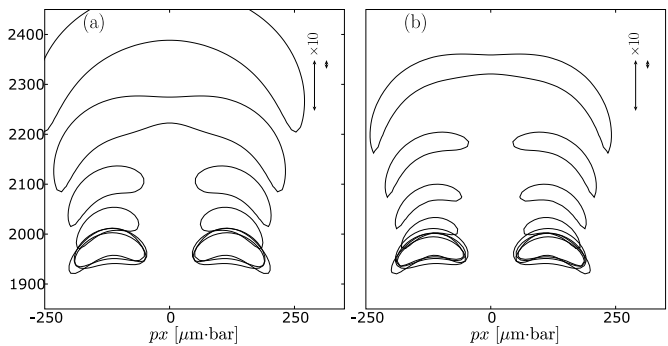}
\caption{\label{positive}
  Evolution of the space charge layers of two
  adjacent positive streamers in air at atmospheric pressure
  at $1 \units{bar}$ (a) and $50 \units{bar}$ (b).
}
\end{figure}

Figure~\ref{positive} shows that the interaction between two positive streamers is qualitatively similar to that
of two negative streamers, since the same two competing phenomena are present.
Initial and boundary conditions are the same, and panels (a) and (b) show positive streamers in air
at atmospheric pressure and at $50 \units{bar}$. The positive streamers take longer to start but
after a relatively short time they merge.

Due to the nonlinear and nonlocal nature of streamer interactions and to the many dimensions of our parameter 
space, it is difficult to develop a simple prediction for streamer merging.  Nevertheless, the following
remarks can be made.  First, to produce enough ionization between the streamers, at least one of the absorption 
lengths of photo-ionization must be larger than or comparable to the streamer distance.  
Second, this photo-ionization is only amplified to a level comparable to that on the streamer head if the field 
is enhanced between the streamers; hence the streamer distance should not be much larger than the streamer radius.  
Finally, an approximation for the relative photo-ionization level can be extracted from \parref{Sph_int_dim} 
as the maximum of the instantaneous rate of photo-ionization on the middle axis produced per impact 
ionization event in the tip of a given streamer, namely $\beta = \xi A(p) h(pd/2)/\pi p^2d^2$. 
Preliminary results show that this parameter influences 
the electron densities in the axis at early stages of the evolution.
We have not found a deterministic law but coalescence is favoured for large $\beta$ and occurs always if $\beta \gtrsim 10^{-11} \units{\mu m^{-3} 
\cdot bar^{-3}}$.

We have developed a code to study the interaction of streamers in full 3D space, 
and we have studied the basic processes that govern the interaction 
of two adjacent streamers.  To focus on the underlying physics of photo-ionization 
we have neglected electron attachment and nontrivial electrode geometries.
Further steps are needed to successfully predict the outcome of experiments and observations;
we mention needle electrodes and the non-alignment of the streamer heads.  
But our results show that for a given pressure $p$, 
electric field $E_0$, oxygen-nitrogen ratio $\eta$
and initial seeds, there is a threshold distance $d^\star$ below which two
streamers coalesce.  This distance, which is experimentally accessible, would
be an indirect measure of the frequency of photoionizing events.
Hence we believe that fully three-dimensional calculations of
streamer dynamics will provide
a suitable test-case for streamer models as they predict easily observable
behavior like attraction, repulsion or branching.

A.L. acknowledges support by Netherland's STW project 06501. U.E. and W.H. acknowledge 
support by the Dutch national program BSIK, in the ICT project BRICKS, 
theme MSV1.


\begin{thebibliography}{99}
\expandafter\ifx\csname natexlab\endcsname\relax\def\natexlab#1{#1}\fi
\expandafter\ifx\csname bibnamefont\endcsname\relax
  \def\bibnamefont#1{#1}\fi
\expandafter\ifx\csname bibfnamefont\endcsname\relax
  \def\bibfnamefont#1{#1}\fi
\expandafter\ifx\csname citenamefont\endcsname\relax
  \def\citenamefont#1{#1}\fi
\expandafter\ifx\csname url\endcsname\relax
  \def\url#1{\texttt{#1}}\fi
\expandafter\ifx\csname urlprefix\endcsname\relax\def\urlprefix{URL }\fi
\providecommand{\bibinfo}[2]{#2}
\providecommand{\eprint}[2][]{\url{#2}}

\bibitem{Raizer} \bibinfo{author}{\bibfnamefont{Y.~P.} \bibnamefont{Raizer}},
  \emph{\bibinfo{title}{Gas Discharge Physics}} (\bibinfo{publisher}{Springer},
  \bibinfo{address}{Berlin}, \bibinfo{year}{1991}).

\bibitem{PSST06} U. Ebert {\it et al.}, Plasma Sour. Sci. Tech. {\bf 15}, S118 (2006).

\bibitem[{\citenamefont{{Stenbaek-Nielsen}
  et~al.}(2007)\citenamefont{{Stenbaek-Nielsen}, {McHarg}, {Kanmae}, and
  {Sentman}}}]{NielsenGRL07}
\bibinfo{author}{\bibfnamefont{H.~C.} \bibnamefont{{Stenbaek-Nielsen}}} {\it et al.},
  \bibinfo{journal}{Geophys. Res. Lett.} \textbf{\bibinfo{volume}{34}},
  \bibinfo{pages}{11105} (\bibinfo{year}{2007}).


\bibitem[{\citenamefont{Pasko et~al.}(1998)\citenamefont{Pasko, Inan, and
  Bell}}]{PaskoGRL98}
\bibinfo{author}{\bibfnamefont{V.}~\bibnamefont{Pasko} {\it et al.}},
  \bibinfo{journal}{Geophys. Res. Lett.} \textbf{\bibinfo{volume}{25}},
  \bibinfo{pages}{2123} (\bibinfo{year}{1998}).

\bibitem[{\citenamefont{Pasko}(2006)}]{PaskoBook}
\bibinfo{author}{\bibfnamefont{V.~P.} \bibnamefont{Pasko}}, in
  \emph{\bibinfo{booktitle}{Sprites, Elves and Intense Lightning Discharges}},
  ed.: M.~F\"ullekrug {\it et al.}, (\bibinfo{publisher}{Springer Netherlands},
  \bibinfo{year}{2006}), pp. \bibinfo{pages}{253--311}.

\bibitem{fra1990} R.~Franz {\it et al.}, Science {\bf 249}, 48 {1990}.

\bibitem{Rodin2004} P.~{Rodin}, I.~{Grekhov},
  \bibinfo{journal}{App. Phys. Lett.} \textbf{\bibinfo{volume}{86}},
  \bibinfo{eid}{243504} (\bibinfo{year}{2005}).

\bibitem[{\citenamefont{{Massala} and {Lesaint}}(2001)}]{Massala01}
\bibinfo{author}{\bibfnamefont{G.}~\bibnamefont{{Massala}}} \bibnamefont{and}
  \bibinfo{author}{\bibfnamefont{O.}~\bibnamefont{{Lesaint}}},
  \bibinfo{journal}{J. Phys. D} \textbf{\bibinfo{volume}{34}},
  \bibinfo{pages}{1525} (\bibinfo{year}{2001}).

\bibitem[{\citenamefont{Array{\'a}s et~al.}(2002)\citenamefont{Array{\'a}s,
  Ebert, and Hundsdorfer}}]{arr2002}
\bibinfo{author}{\bibfnamefont{M.}~\bibnamefont{Array{\'a}s} {\it et al.}},
  \bibinfo{journal}{Phys.\ Rev.\ Lett.} \textbf{\bibinfo{volume}{88}},
  \bibinfo{pages}{174502} (\bibinfo{year}{2002}).

\bibitem[{\citenamefont{Meulenbroek et~al.}(2005)\citenamefont{Meulenbroek,
  Ebert, and Sch{\"a}fer}}]{meu2005}
\bibinfo{author}{\bibfnamefont{B.}~\bibnamefont{Meulenbroek} {\it et al.}},
  \bibinfo{journal}{Phys.\ Rev.\ Lett.} \textbf{\bibinfo{volume}{95}},
  \bibinfo{pages}{195004} (\bibinfo{year}{2005}).

\bibitem{Briels2006} T.M.P. Briels {\it et al.}, J. Phys. D {\bf 39}, 5201 (2006).

\bibitem[{\citenamefont{Gerken et~al.}(2000)\citenamefont{Gerken, Inan, and
  Barrington-Leigh}}]{ger2000}
\bibinfo{author}{\bibfnamefont{E.}~\bibnamefont{Gerken} {\it et al.}},
  \bibinfo{journal}{Geophys. Res. Lett.} \textbf{\bibinfo{volume}{27}},
  \bibinfo{pages}{2637} (\bibinfo{year}{2000}).

\bibitem{Cummer06} 
\bibinfo{author}{\bibfnamefont{S.~A.} \bibnamefont{{Cummer}}} {\it et al.},
\bibinfo{journal}{Geophys. Res. Lett.}
  \textbf{\bibinfo{volume}{33}}, \bibinfo{pages}{4104} (\bibinfo{year}{2006}).
 

\bibitem[{\citenamefont{Winands et~al.}(2006)\citenamefont{Winands, Z., Pemen,
  van Heesch~E.J.M., Yan, and van Veldhuizen}}]{Winands06}
\bibinfo{author}{\bibfnamefont{G.}~\bibnamefont{Winands}}, {\it et al.},
  \bibinfo{journal}{J. Phys. D} \textbf{\bibinfo{volume}{39}},
  \bibinfo{pages}{3010} (\bibinfo{year}{2006}).

\bibitem{Akyuz03} 
\bibinfo{author}{\bibfnamefont{M.}~\bibnamefont{Akyuz}}, {\it et al.},
  \bibinfo{journal}{J. Electrostatics} \textbf{\bibinfo{volume}{59}},
  \bibinfo{pages}{115} (\bibinfo{year}{2003}).


\bibitem[{\citenamefont{Naidis}(1996)}]{Naidis96}
\bibinfo{author}{\bibfnamefont{G.~V.} \bibnamefont{Naidis}},
  \bibinfo{journal}{J. Phys. D: Appl. Phys.}
  \textbf{\bibinfo{volume}{29}}, \bibinfo{pages}{779} (\bibinfo{year}{1996}).

\bibitem{FingerLetter} A.~Luque, F. Brau, U. Ebert, arXiv:0708.1722.

\bibitem[{\citenamefont{Kulikovsky}(1998)}]{Kulikovsky98b}
\bibinfo{author}{\bibfnamefont{A.~A.} \bibnamefont{Kulikovsky}},
  \bibinfo{journal}{Phys. Lett. A} \textbf{\bibinfo{volume}{245}},
  \bibinfo{pages}{445} (\bibinfo{year}{1998}).

\bibitem[{\citenamefont{Pancheshnyi}(2005)}]{Pancheshnyi05b}
\bibinfo{author}{\bibfnamefont{S.}~\bibnamefont{Pancheshnyi}},
  \bibinfo{journal}{Plasma~Sour.~Sci.~Tech.} \textbf{\bibinfo{volume}{14}},
  \bibinfo{pages}{645} (\bibinfo{year}{2005}).

\bibitem[{\citenamefont{Liu and Pasko}(2006)}]{LiuPasko06}
\bibinfo{author}{\bibfnamefont{N.}~\bibnamefont{Liu,}}
  \bibinfo{author}{\bibfnamefont{V.~P.} \bibnamefont{Pasko}},
  \bibinfo{journal}{J.~Phys.~D} \textbf{\bibinfo{volume}{39}},
  \bibinfo{pages}{327} (\bibinfo{year}{2006}).

\bibitem[{\citenamefont{S\'{e}gur et~al.}(2006)\citenamefont{S\'{e}gur,
  Bourdon, Marode, Bessieres, and Paillol}}]{Segur06}
\bibinfo{author}{\bibfnamefont{P.}~\bibnamefont{S\'{e}gur}} {\it et al.},
\bibinfo{journal}{Plasma Sour. Sci.
  Tech.} \textbf{\bibinfo{volume}{15}}, \bibinfo{pages}{648}
  (\bibinfo{year}{2006}).

\bibitem[{\citenamefont{Luque et~al.}(2007)\citenamefont{Luque, Ebert, Montijn,
  and Hundsdorfer}}]{Luque07}
\bibinfo{author}{\bibfnamefont{A.}~\bibnamefont{Luque} {\it et al.}},
  \bibinfo{journal}{App.~Phys.~Lett.} \textbf{\bibinfo{volume}{90}},
  \bibinfo{eid}{081501} (\bibinfo{year}{2007}).

\bibitem[{\citenamefont{Bourdon et~al.}(2007)\citenamefont{Bourdon, Pasko, Liu,
  C\'{e}lestin, S\'{e}gur, and Marode}}]{Bourdon07}
\bibinfo{author}{\bibfnamefont{A.}~\bibnamefont{Bourdon}} {\it et al.,}
  \bibinfo{journal}{Plasma Sour. Sci. Tech.} \textbf{\bibinfo{volume}{16}},
  \bibinfo{pages}{656} (\bibinfo{year}{2007}).

\bibitem[{\citenamefont{Penney and Hummert}(1970)}]{pen1970}
\bibinfo{author}{\bibfnamefont{G.}~\bibnamefont{Penney}},
  \bibinfo{author}{\bibfnamefont{G.}~\bibnamefont{Hummert}},
  \bibinfo{journal}{J. Appl. Phys} \textbf{\bibinfo{volume}{41}},
  \bibinfo{pages}{572} (\bibinfo{year}{1970}).

\bibitem[{\citenamefont{Vitello et~al.}(1994)\citenamefont{Vitello, Penetrante,
  and Bardsley}}]{vit1994}
\bibinfo{author}{\bibfnamefont{P.}~\bibnamefont{Vitello} {\it et al.}},
  \bibinfo{journal}{Phys.\ Rev.~E} \textbf{\bibinfo{volume}{49}},
  \bibinfo{pages}{5574} (\bibinfo{year}{1994}).

\bibitem[{\citenamefont{Montijn et~al.}(2006)\citenamefont{Montijn,
  Hundsdorfer, and Ebert}}]{Montijn06}
\bibinfo{author}{\bibfnamefont{C.}~\bibnamefont{Montijn} {\it et al.}},
  \bibinfo{journal}{J. Comput. Phys.} \textbf{\bibinfo{volume}{219}},
  \bibinfo{pages}{801} (\bibinfo{year}{2006}).

\bibitem[{\citenamefont{Zheleznyak et~al.}(1982)\citenamefont{Zheleznyak,
  Mnatsakanyan, and Sizykh}}]{zhe1982}
\bibinfo{author}{\bibfnamefont{M.}~\bibnamefont{Zheleznyak} {\it et al.}},
  \bibinfo{journal}{High Temp.} \textbf{\bibinfo{volume}{20}},
  \bibinfo{pages}{357} (\bibinfo{year}{1982}).

\bibitem[{\citenamefont{Wackers}(2005)}]{wac2005}
\bibinfo{author}{\bibfnamefont{J.}~\bibnamefont{Wackers}}, \bibinfo{journal}{J.
  Comp. Appl. Math.} \textbf{\bibinfo{volume}{180}}, \bibinfo{pages}{1}
  (\bibinfo{year}{2005}).

\bibitem{TanjaSca08} T.M.P.~Briels, E.M. van Veldhuizen, U. Ebert, 
  arXiv:0805.1364.

\bibitem[{\citenamefont{Kulikovsky}(2000)}]{kul2000-2}
\bibinfo{author}{\bibfnamefont{A.}~\bibnamefont{Kulikovsky}},
  \bibinfo{journal}{J.~Phys.~D} \textbf{\bibinfo{volume}{33}},
  \bibinfo{pages}{1514} (\bibinfo{year}{2000}).

\end{thebibliography}
\end{document}